\documentclass[11pt]{article}
\usepackage{graphicx}
\usepackage{graphics}
\usepackage{epsfig}
\usepackage{amsmath}
\usepackage[dvips]{feynmp}
\newif\ifdimspec
\def\figsize#1{\dimspecfalse \checkdim#1\end
\ifdimspec
  \def\figureWidth{#1}%
\else
  \def\figureWidth{#1 in}\fi}
\def\checkdim#1{\ifx#1\end \let\next=\relax
  \else \ifcat#1a \dimspectrue \fi \let\next=\checkdim\fi \next}
\newcommand{\lblcaption  }[2]{\caption{#2\label{\secname#1}}}
\newcommand{\twoAngFiguresEPS}[6]
{
\begin{figure}
\begin{center}
\figsize{#4}
\begin{minipage}[t]{3.2in}
\begin{center}
\epsfig{file=\sectiondir/#1.eps,width=\figureWidth,angle=#5}
\end{center}
\end{minipage}
\begin{minipage}[t]{3.2in}
\begin{center}
\epsfig{file=\sectiondir/#2.eps,width=\figureWidth,angle=#6}
\end{minipage}
\end{center}
\lblcaption{#1}{#3}
\end{figure}
}
\newcommand{\BABARPubYear}    {01}

\newcommand{\BABARConfNumber} {28}
\newcommand{\SLACPubNumber} {8982}

\input pubboard/babarsym
\def\BRbztodstdstNum {\ensuremath{(8.0 \pm 1.6(stat) \pm
1.2(syst))\times 10^{-4}}}
\def\BRbztodstdst {\ensuremath{{\BR}(\Bztodstdst) = \BRbztodstdstNum}}

\def\fsideVal {1.72 \xspace}
\def\fsideErr {0.10 \xspace}

\def\TotLumi {23.3\invfb}
\def\OnResLumi {20.7\invfb}

\def\NBB {(22.7 \pm 0.36) \times 10^{6}}

\def\DelELowSide {50\mev}
\def\DelEHiSide {200\mev}
\def\mesLowSide {5.20\gevcc}
\def\mesMidSide {5.26\gevcc}
\def\mesHiSide  {5.29\gevcc}

\def\Dstarp     {\ensuremath{D^{*+}}\xspace}
\def\Dstarm     {\ensuremath{D^{*-}}\xspace}
\def\Dstarpm    {\ensuremath{D^{*\pm}}\xspace}
\def\Dp         {\ensuremath{D^+}\xspace}
\def\Dm         {\ensuremath{D^-}\xspace}

\def\BtoDDbar	{\ensuremath{B \to D^{(*)} \Dbar^{(*)}}\xspace}

\def\BtoDsDbar	{\ensuremath{B \to D^{(*)}_s \Dbar^{(*)}}\xspace}

\def\BztoDDbar	{\ensuremath{B^0 \to D^{(*)+} D^{(*)-}}\xspace}

\def\Bpmtodstdstz {\ensuremath{B^{+} \to D^{*+} D^{*0}}\xspace}
\def\Bztodstdcc {\ensuremath{B^{0} \to D^{*+} D^{-}}\xspace}
\def\DstDst {\ensuremath{D^{*+} D^{*-}\xspace}}
\def\Dzpip{\ensuremath{D^0 \pi^+ \xspace}}
\def\Dzpim{\ensuremath{D^0 \pi^- \xspace}}
\def\Dppiz{\ensuremath{D^+ \pi^0 \xspace}}
\def\Dmpiz{\ensuremath{D^- \pi^0 \xspace}}

\def\Dstptopip  {\ensuremath{\Dstarp \to \Dz\pip}\xspace}
\def\Dstptopiz  {\ensuremath{\Dstarp \to \Dp\piz}\xspace}
\def\Dstztopiz  {\ensuremath{\Dstarz \to \Dz\piz}\xspace}
\def\Dstztogam  {\ensuremath{\Dstarz \to \Dz\gamma}\xspace}

\def\DeltaEStd  {\ensuremath{\Delta E} \xspace}
\def\SsqovSpB   {\ensuremath{S^2/(S+B)}\xspace}

\def\chisqM     {\ensuremath{\chi^2_{Mass}}\xspace}

\long\def\inst#1{\par\nobreak\kern 4pt\nobreak
    {\it #1}\par\vskip 10pt plus 3pt minus 3pt}

\begin{document}

\pagestyle{empty}

%\begin{flushleft}
%\babar\ Analysis Document \#291, Version 7 \\
%\end{flushleft}

\begin{flushright}
\babar-CONF-\BABARPubYear/\BABARConfNumber \\
%\babar-PUB-\BABARPubYear/\BABARPubNumber \\
SLAC-PUB-\SLACPubNumber \\
September, 2001 \\

\end{flushright}

\par\vskip 4cm

% Title of the paper
\begin{center}
\Large \bf Study of \BtoDDbar decays with the \babar\ detector 
\end{center}
\bigskip

\begin{center}
\large The \babar\ Collaboration\\
\mbox{ }\\
\today
\end{center}
\bigskip \bigskip

% Abstract
\begin{center}
\large \bf Abstract
\end{center}
Decays of the type \BtoDDbar\ can be used to 
provide a measurement of the parameter
\stwob of the Unitarity Triangle that is complementary to that
derived from the mode \bpsiks.  Here we report a
measurement of the branching fraction 
and a study of the \CP parity content
for the decay \Bztodstdst\ with the
\babar\ detector.  With data corresponding to an integrated luminosity 
of \OnResLumi\ collected at the \FourS\ resonance
during 1999-2000, 
we determine the branching fraction to be
\BRbztodstdst. 
The measured fraction of the component with odd \CP
parity is $0.22 \pm 0.18(stat) \pm 0.03(syst)$.
Observation of a significant number of candidates in the decay modes
\Bztodstd and \Bpmtodstdstz is reported. All results presented in this note
are preliminary.

\vfill
\begin{center}
Submitted to the 9$^{th}$ International Symposium On Heavy Flavor Physics, \\
9/10---9/13/2001, Pasadena, CA, USA
\end{center}

\vspace{1.0cm}
\begin{center}
{\em Stanford Linear Accelerator Center, Stanford University, 
Stanford, CA 94309} \\ \vspace{0.1cm}\hrule\vspace{0.1cm}
Work supported in part by Department of Energy contract DE-AC03-76SF00515.
\end{center}

\newpage
\pagestyle{plain}

% Input author list file
\begin{center}
\small

The \babar\ Collaboration,
\bigskip

%% author list as of 22-Aug-2001 (584 authors)
B.~Aubert,
D.~Boutigny,
J.-M.~Gaillard,
A.~Hicheur,
Y.~Karyotakis,
J.~P.~Lees,
P.~Robbe,
V.~Tisserand
\inst{Laboratoire de Physique des Particules, F-74941 Annecy-le-Vieux, France }
A.~Palano,
A.~Pompili
\inst{Universit\`a di Bari, Dipartimento di Fisica and INFN, I-70126 Bari, Italy }
% ...

\end{center}\newpage

% reset footnote counter
\setcounter{footnote}{0}

% The body of the paper starts here
\section{Introduction}
\label{sec:Introduction}
One of the most important goals of the \babar\ experiment is to
precisely measure the angles of the Unitarity Triangle.  While the
decay \bpsiks\ can be used to measure \stwob, the Standard
Model predicts that the time-dependent \CP-violating asymmetries
in the decays \BztoDDbar can also be used to measure the same quantity~\cite{ref:cc}.  
An independent measurement of \stwob in \BztoDDbar modes is especially important 
since several typical extensions to the Standard Model can lead to different assymmetries 
between this quantity in \BztoDDbar\ events and charmonium events~\cite{GWF}.
Measurements of \stwob in \BztoDDbar would thus provide stringent tests of \CP-violation in the
Standard Model and have significant potential to indicate deviations from it.
However, the vector-vector decay \Bztodstdst is not a pure \CP
eigenstate and a sizeable dilution of the measured asymmetry can be
produced by a non-negligible $P$-wave \CP-odd component.  The
dilution can, in principle, be completely removed by a time-dependent angular
analysis of the decay products~\cite{ref:dunietz}.
\par The rate for the Cabibbo-suppressed decays \BtoDDbar can be estimated
from the measured rate of the Cabibbo-favored decays \BtoDsDbar:
\begin{equation}
{\BR}(\BtoDDbar) \approx
\left(\frac{f_{D^{(*)}}}{f_{D_s^{(*)}}}\right) \tan^2\theta_C
\cdot{\BR}(\BtoDsDbar),
\end{equation}
where $\theta_C$ is the Cabibbo angle, and $f_{D^{(*)}}$ and
$f_{D_S^{(*)}}$ are decay constants.
From this it follows that the \BtoDDbar branching fractions are of the
order of $10^{-3}$. Previous measurements of branching fractions and upper 
limits for these modes are summarized in Table \ref{prevbr}. 
\par In section~\ref{sec:babar} we describe briefly the \babar\ detector and the dataset.
Section~\ref{sec:Analysis} describes the measurement of the branching fraction \Bztodstdst\ 
and section~\ref{sec:Transversity} the corresponding angular analysis to extract the 
CP-even component of the decay. Finally, in section~\ref{sec:Observation} we discuss the 
observation of the decays \Bztodstdcc\ and \Bpmtodstdstz. Section~\ref{sec:Summary} summarizes our results.

\begin{table}[bh]
\caption{\label{prevbr}Summary of branching fraction and upper limit
measurements performed by the CLEO \cite{cleoprd62} and ALEPH 
\cite{alephepj98} experiments. Upper limits are quoted at the 90\% confidence level.}

\begin{center} \begin{tabular}{llc}
\hline \hline
\multicolumn{1}{c}{Decay} &  & Branching Fraction ($\times 10^{-4}$) \\[0.5ex]
\hline
\Bztodstdst & \cite{cleoprd62} &  $9.9^{+4.2}_{-3.3}(stat)\pm 1.2(syst)$ \\
\Bztodstdcc   & \cite{cleoprd62} &  $<6.3$ \\
\Bpmtodstdstz & \cite{alephepj98}&  $<110.$ \\
\hline \hline
\end{tabular} \end{center}
\end{table}

\section{The \babar\ detector and dataset}
\label{sec:babar}
The data used in this analysis were collected with the \babar\
detector~\cite{ref:babar} at the \pep2\ storage ring~\cite{ref:pepii}
located at the Stanford Linear Accelerator Center.  
%The results presented in this paper are based on data taken during the 1999-2000 run.  
This data sample represents an integrated
luminosity of \TotLumi, with \OnResLumi collected on the \FourS
resonance.  The total number of \BB pairs produced in this sample was
$N_{\BB} = \NBB$.

Charged particles are detected and their momenta measured with the
combination of a 40-layer drift chamber (DCH) and a five-layer silicon
vertex tracker (SVT) embedded in a 1.5\,T solenoidal magnetic field.
Photons are detected by a CsI electromagnetic calorimeter (EMC) that
provides excellent angular and energy resolutions with a high
efficiency for energies above 20\mev.  Charged particle
identification is provided by the specific ionization loss (\dedx)
in the tracking devices and by an internally reflecting ring-imaging
Cherenkov detector (DIRC) covering the barrel region of the detector.

\section{Measurement of the \Bztodstdst branching Fraction}
\label{sec:Analysis}

\Bz mesons are exclusively reconstructed by combining two charged \Dstar
candidates reconstructed in a number of \Dstar and $D$ decay modes.
Events are pre-selected by requiring that there be
three or more charged tracks and that the normalized second Fox-Wolfram
moment~\cite{ref:fox} of the event be less than 0.6.  We also require
that the cosine of the angle between the reconstructed $B$ direction
and the thrust axis of the rest of the event be less than 0.9.

Charged kaon candidates are required to be inconsistent with the pion
hypothesis, as inferred from the Cherenkov angle
measured by the DIRC and the specific ionization measured by the SVT and DCH.  
No particle identification requirements are made
for the kaon from the decay $\Dz \to \Km \pip$.

$\KS \to \pip\pim$ candidates are required to have an invariant mass
within 25\mevcc of the nominal \KS mass~\cite{pdg}.  The opening angle between the
flight direction and the momentum vector of the \KS candidate is
required to be less than 200\mrad, and the transverse flight distance
from the primary event vertex must be greater than 2\,mm.

Neutral pion candidates are formed from
pairs of photons in the EMC with energy above 30\mev,
an invariant mass within 20\mevcc\ of the nominal \piz\ mass,
and a summed energy greater than 200\mev.
A mass-constraint fit is then applied to these \piz\ candidates. 
The \piz from \Dstptopiz decays (``soft'' \piz),
however, is required to have an invariant mass within
35\mevcc\ of the nominal \piz\ mass and momentum in the \FourS frame
of $70 < p^* < 450\mevc$, with no requirement on the summed
photon energies.

The decay modes of the \Dz and \Dp used in 
this analysis were selected
by an optimization of \SsqovSpB based on Monte Carlo simulations,
where $S$ is the expected number of signal events and $B$ is the
expected number of background events.  
The \Dz and \Dp modes used and their branching fractions are
summarized in Table~\ref{dzbr}.  \Dz (\Dp) meson candidates are required to
have an invariant mass within 20\mevcc of the nominal \Dz (\Dp) mass.

\begin{table}[ht]
\caption{\label{dzbr} \Dz and \Dp decay modes and branching
fractions~\cite{pdg}.  The branching fraction for $\KS \to \pipi$ is
included for modes containing a \KS.}
\begin{center} \begin{tabular}{lc}
\\ \hline \hline 
Decay Mode & Branching Fraction (\%) \\[0.5ex]
\hline
$\Dz \to \Km \pip$           & $3.83 \pm 0.09$ \\
$\Dz \to \Km \pip \piz$      & $13.9 \pm 0.9$  \\
$\Dz \to \Km \pip \pip \pim$ & $7.49 \pm 0.31$ \\
$\Dz \to \KS \pip \pim$      & $1.85 \pm 0.14$ \\
\hline
Total \Dz Branching Fraction & 27.1 \\
\hline \hline
Decay Mode & Branching Fraction (\%) \\[0.5ex]
\hline
$\Dp \to \Km \pip \pip$      & $9.0 \pm 0.6$   \\
$\Dp \to \KS \pip$           & $0.99 \pm 0.09$ \\
$\Dp \to \Km \Kp \pip$       & $0.87 \pm 0.07$ \\
\hline
Total \Dp Branching Fraction & 10.9 \\
\hline \hline
\end{tabular} \end{center}
\end{table}

The \Dstarp mesons
are reconstructed in their decays \Dstptopip and \Dstptopiz. 
We include for this analysis the decay combinations \DstDst decaying 
to (\Dzpip, \Dzpim) or (\Dzpip, \Dmpiz), but not (\Dppiz,\Dmpiz) due to 
the smaller branching fraction and larger expected backgrounds.
The branching fractions for these modes are summarized in Table \ref{dstarbr}.
\Dz and \Dp candidates are subjected to a mass-constraint fit and
then combined with soft pion candidates.  A vertex fit is performed
that includes the position of the beam spot to improve the angular
resolution of the soft pion.

\begin{table}[t]
\caption{\label{dstarbr}\Dstar and \Dstarz decay modes and branching fractions~\cite{pdg}.
\Dstarz is used for the \Bpmtodstdstz analysis described in Section~\ref{sec:Observation}.}

\begin{center} \begin{tabular}{lcc}
\hline \hline
Particle & Decay Mode & Branching Fraction (\%) \\[0.5ex]
\hline
\Dstarp & \Dstptopip & $67.7 \pm 0.5$ \\
        & \Dstptopiz & $30.7 \pm 0.5$ \\
Total Visible \Dstarp Branching Fraction & & 98.4 \\
\hline \hline
\Dstarz & \Dstztopiz & $61.9 \pm 2.9$ \\ 
        & \Dstztogam & $38.1 \pm 2.9$ \\
Total \Dstarz Branching Fraction & & 100.0 \\
%\hline
%\Dstarz & \Dstztopiz & $61.9 \pm 2.9$ \\
\hline \hline
\end{tabular} \end{center}
\end{table}

To select \Bz candidates with well reconstructed \Dstar and $D$
mesons, we construct a $\chi^2$ that includes all measured \Dstar
and $D$ masses:

\begin{eqnarray*}
\chisqM =& 
   \left(\displaystyle \frac{m_D - m_{D_{PDG}}}{\sigma_{m_D}}\right)^2
 + \left(\displaystyle \frac{m_{\Db} - m_{\Db_{PDG}}}{\sigma_{m_{\Db}}}\right)^2 \\
  &\quad + \left(\displaystyle\frac{\Delta m_{D^{*}} - \Delta
   m_{\Dstar_{PDG}}}{\sigma_{\Delta m_{D^{*}}}}\right)^2
 + \left(\displaystyle\frac{\Delta m_{\overline{D}^*} - \Delta
   m_{\Dstar_{PDG}}}{\sigma_{\Delta m_{D^{*}}}}\right)^2
\end{eqnarray*}
where the subscript $PDG$ refers to the nominal value, and $\Delta m_{D^{*}}$ 
is the $\Dstar - D$ mass difference.  For $\sigma_{m_D}$ we use
values computed for each $D$ candidate, while for $\sigma_{\Delta m_{D^{*}}}$
we use fixed values of 0.83\mevcc for \Dstptopip and 1.18\mevcc
for \Dstptopiz.  A requirement that $\chisqM < 20$ is applied to all \Bz
candidates.  In events with more than one \Bz candidate, we choose the
candidate with the lowest value of \chisqM.

A $B$ meson candidate is characterized by two kinematic variables.  We
use the energy-substituted mass, \mes, defined as
$$\mes \equiv \sqrt{{E_{Beam}^{*2}} - {p_B^*}^2}$$
and the difference of the $B$ candidate's energy from the beam energy,
\DeltaEStd,
$$\DeltaEStd \equiv E_{B}^* - E_{Beam}^* $$
where $E_{B}^*$ ($p_B^*$) are the energy (momentum) of the \B\ candidate
in the center-of-mass frame and $E_{Beam}^*$ is one-half of the
center-of-mass energy.
The signal region in the \DeltaEStd {\it vs.} \mes plane is defined to be
$|\DeltaEStd| < 25\mev$ and $5.273 < \mes < 5.285\gevcc$.
The width of this region corresponds to approximately $\pm 2.5\sigma$
in both \DeltaEStd and \mes.

These values on \chisqM, \mes, and \DeltaEStd were chosen based on
an optimization of $S^2/(S+B)$, where S is the expected number 
of signal events and B is the expected number of background events.
The optimization process was done entirely with
samples of signal and generic \BB and \ccbar Monte Carlo where the 
background distribution is taken from a sideband region, defined as
$$ |\DeltaEStd| < \DelEHiSide $$
$$ \mesLowSide < \mes < \mesMidSide $$
and
$$ \DelELowSide < |\DeltaEStd| < \DelEHiSide $$
$$ 5.26 < \mes < \mesHiSide $$
These values were chosen based on a maximization of $S^2/(S+B)$ with a tendency
towards looser cut values to reduce any possible systematic error incurred
due to the differences in the reconstructed mass resolutions between
data and Monte Carlo.

To determine the number of signal events in the signal 
region, we must estimate the expected contribution from background.
This is done by scaling the number of events seen in the data sideband 
(defined above) with a scaling factor which gives a measure of the
relative areas of the signal region to the sideband region.  
We parameterize the shape of the background in the \DeltaEStd {\it vs.}
\mes plane as the product of an ARGUS
function~\cite{ref:argus} in \mes and a first-order polynomial in \DeltaEStd.
Based on this parameterization we estimate that the ratio
of the number of background events in the signal region
to the number in the sideband region is
$(\fsideVal \pm \fsideErr)\times 10^{-2}$.
The uncertainty is derived from the observed variation of this ratio
under alternative assumptions 
for the background shape in \mes\ and \DeltaEStd.

\begin{figure}[!htb]
\begin{center}
\includegraphics[height=7cm]{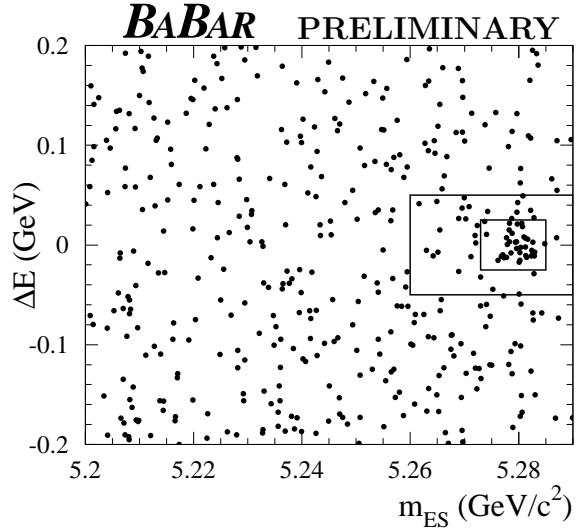}
\put(-105,185){{\large \bf PRELIMINARY}}
\caption{Distribution of \Bztodstdst\ events in the \DeltaEStd {\it vs.} \mes plane.
The small box indicates the signal region, while the sideband region
is everything outside the larger box.
}
\label{fig:mesdeltae}

\end{center}
\end{figure}

\begin{figure}[!htb]
\begin{center}
\includegraphics[height=7cm]{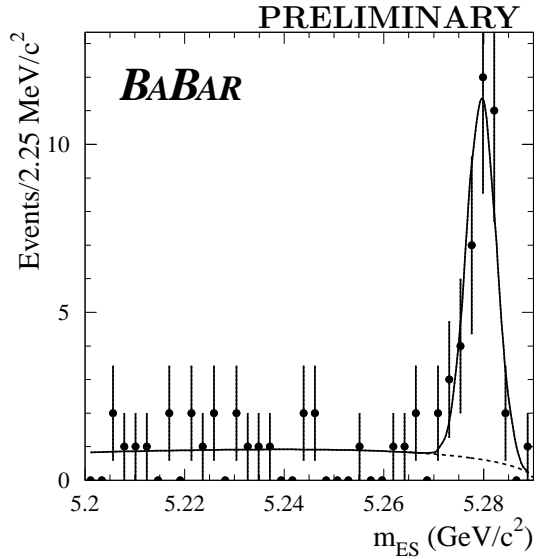}
\put(-105,200){{\large \bf PRELIMINARY}}
\caption{Distribution of \Bztodstdst\ events in \mes plane with a cut of $|\DeltaEStd|
< 25\,\mev$ applied.  The curve represents a fit to the distribution
of the sum of a Gaussian to model the signal and an ARGUS 
function~\cite{ref:argus} to model the background shape.
}
\label{fig:mes}
\end{center}
\end{figure}

Figure~\ref{fig:mesdeltae} shows the events in the \DeltaEStd {\it vs.}
\mes plane after all selection criteria have been applied.  The small
box in the figure indicates the signal region defined above, and the
sideband is the entire plane excluding the region bounded by the
larger box outside the signal region.  There are a total of 38 events
located in the signal region, with 363 events in the sideband region.
The latter, together with the effective ratio of areas of the signal
region to the sideband region, implies an expected number of
background events in the signal region of $6.24 \pm 0.33(stat) \pm
0.36 (syst)$.
The quoted systematic
uncertainty comes from the
background shape variation discussed previously.  Figure~\ref{fig:mes}
shows a projection of the data on the \mes axis after requiring
$|\DeltaEStd| < 25\mev$.
 
We use a detailed Monte Carlo simulation of the \babar\ detector to
determine the efficiency for reconstructing the signal.  This,
together with the total number of \BB pairs produced during data
collection, allows us to determine a preliminary branching fraction for
\Bztodstdst to be

$$\BRbztodstdst$$

The dominant systematic uncertainty in this measurement comes from our
level of understanding of the charged particle tracking efficiency (9.4\%).
The high charged particle multiplicity in this decay mode makes this
measurement particularly sensitive to tracking efficiency. Systematic errors 
were assigned on a per track basis for $\pi$, $K$ and slow $\pi$,
and were added linearly due to large correlations.  The imprecisely
known partial-wave content of the \Bztodstdst\ final state
is another source of systematic biases. 
Monte Carlo events in each of the two extremes of transversity amplitudes 
$(A_{//},\sqrt{2}A_{0},A_{\perp})=(1.,0.,0.)$ and $(0.,1.,0.)$
were generated and reconstructed~\cite{transamplitudes}. 
Although both mixtures correspond to $R_t=0$, the resulting $p_t$ 
distributions of the slow pion
represent the two extreme cases of possible $p_t$ distributions. 
The change in the reconstruction efficiency of  
these final angular states is quoted as systematic error (6.6\%). 
Other significant systematic biases arise due to the
uncertainties on the ${\Dstar}^+$, \Dz and
$D^+$ branching fractions (5.6\%) and the differences in mass
resolutions between Monte Carlo and data (4.1\%). 
Possible contributions from peaking backgrounds was found to
be negligable.The total systematic uncertainty from all sources is 14.5\%.

%Give the principle of the analysis, the strengths and weaknesses of
%\babar\ in confronting this problem and the basic event selection criteria.

%Describe the analysis is a clear step-by-step manner.
%Refer in the text to all figures and tables, such as Figs.~\ref{fig:pawfigure}
%and \ref{fig:rootfigure}
%and Table~\ref{tab:example}, in the order in which they appear.
%
%\begin{figure}[!htb]
%\begin{center}
%\includegraphics[height=7cm]{pubboard/pawfigure.eps}
%\caption{This was made using pawfigure.kumac, which calls the
%plot.kumac and babar.kumac paw macros.
%Define all symbols used in the figure in its caption.
%Note the caption goes below the figure.}
%\label{fig:pawfigure}
%\end{center}
%\end{figure}
%

\section{Determination of the $CP$ content of \Bztodstdst}
\label{sec:Transversity}
%Clearly present the analysis of systematic uncertainties.
The fraction of the $CP$-odd component, $R_t$, of the decay \Bztodstdst
can be determined from the angular distribution in the transversity
basis~\cite{ref:dunietz}:
\begin{equation}
\frac{1}{\Gamma}\frac{\rm{d}\Gamma}{\rm{d}\cos\theta_{tr}} = 
\frac{3}{4}(1-R_t)\sin^2\theta_{tr} + 
\frac{3}{2}R_t\cos^2\theta_{tr}
\label{angdis}
\end{equation}
Here $\Gamma$ is the decay rate and 
$\theta_{tr}$ is the polar angle defined as the angle
between the normal to the ${\Dstar}^-$ decay plane and the \pip line
of flight in the ${\Dstar}^+$ rest frame.

We perform an unbinned maximum likelihood fit to the 38 events in the
signal region described in the previous section.  The fit takes into
account the presence of background, whose properties are derived from
the sideband sample, and the angular resolution $\sigma_{\theta_{tr}}$
estimated from the simulated data samples.
We define the likelihood function to be
\begin{equation}
\begin{array}{r@{}c@{}l}
\displaystyle
{\cal{L}}=
\prod_{i=1,n}{\cal{L}}_i=
\prod_{i=1,n}
\left[p\rule{0mm}{4mm}\right.
&\times&
\left.{\cal{F}}(\theta_{tr,i},\sigma_{\theta,i},R_t^{sig})+
\right.\\[5mm]
(1-p)
&\times&
\left.{\cal{F}}(\theta_{tr,i},\sigma_{\theta,i},R_t^{bkg})
\rule{0mm}{4mm}\right],
\end{array}
\label{ll1}
\end{equation}
where $n$ is the number of selected events and the contribution to the
total likelihood from the $i$-th event, ${\cal{L}}_i$, is defined 
in terms of the purity $p$ of the sample and the 
probability density functions 
$\left.{\cal{F}}(\theta_{tr,i},\sigma_{\theta,i},R_t)\right.$
for the signal and background: 
$R_t^{sig}$ and $R_t^{bkg}$ are the parameters describing the shapes of the
signal and background angular distributions, respectively, and 
$\theta_{tr,i}$ is the measured transversity angle
in event $i$.
The probability density functions $\left.{\cal{F}}\right.$ 
are obtained from the
convolution of the angular distribution (Eq.~\ref{angdis}) with
Gaussian resolution functions describing the measurement errors
$\sigma_{\theta,i}$.  We note that Eq.~\ref{angdis} is the 
differential decay rate
$\Gamma$ integrated over the full ranges of the other two decay angles in 
the transversity basis; this assumes a flat acceptance and needs to be 
corrected for in the final determination of $R_t$.
An empirical study of the events in the sideband region showed that
the background angular distribution can be described as constant in 
$\cos \theta_{tr}$, as expected. The parameterization in Eq.~\ref{ll1}
can describe deviations from this behaviour and was used to estimate
the background contribution to the systematic uncertainty on $R_t$.

The fit procedure was tested on several samples of \Bztodstdst Monte
Carlo events generated with different $R_t$ values and different
sets of decay amplitudes in order to determine the possible bias induced 
by angular acceptance effects correlated with the inefficiency in detecting
soft pions from $D^{*}$ decays below a threshold in transverse momentum 
of about $70\mevc$.
The fitted $R_t$ values were fully consistent with the generated values
in the limit of negligible soft pion inefficiency, but could be biased, 
depending on the decay amplitudes, when the pion detection threshold 
was taken into account. A correction and a systematic error were estimated
to describe these effects (described below). 

The value of $R_t^{bkg}$ was evaluated by fitting the 363 events in
the sideband region and setting $p=0$ in Eq.~\ref{ll1}.  The result
of this fit was
%\begin{equation} 
$R_t^{bkg}=0.29\pm0.04$, 
%\end{equation}
compatible with the value expected for a flat distribution ($R_t=1/3$). 

\begin{figure}[htb]
\begin{center}
\epsfig{file=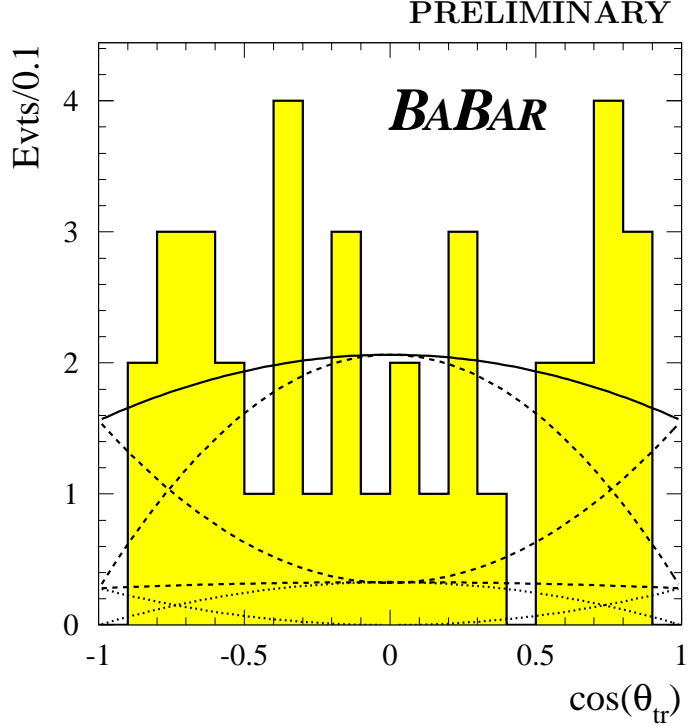,width=9cm}
\put(-105,265){{\large \bf PRELIMINARY}}
\caption{The $\cos\theta_{tr}$ distribution from the unbinned ML fit,
superimposed on the histogram of the \Bztodstdst\ 
candidates in the signal region. The dotted and dashed lines 
represent the fitted \CP\ components for the signal and 
background contributions. The lower two dotted curves represent 
the two background fitted components, 
proportional to $\sin^2 \theta_{tr}$ and 
$\cos^2 \theta_{tr}$ respectively; 
their sum (dashed curve) is almost constant at
about $0.3$ events/bin. Similarly, the two upper dashed curves represent the 
\CP\ components of the signal. 
The solid  line represents the $\cos\theta_{tr}$ 
distribution from the unbinned ML fit for the selected events.}
\label{fig:fit_sig}
\end{center}
\end{figure}

To determine $R_t^{sig}$, we fit the 38 events in the signal region
with $R_t^{bkg}$ fixed to 0.29 and $p$ fixed at 83.6\%.  The result of
the fit to the signal region, without correction for angular acceptance bias, is 
%\begin{equation}
$R_t^{sig}=0.25\pm0.18(stat)$,
%\end{equation}
and is shown in Figure~\ref{fig:fit_sig}.  The fit was repeated
with both $R_t^{sig}$ and $R_t^{bkg}$ floating, giving the same central values 
with a rather small correlation ($-0.04$) between the parameters.

\begin{figure}[htb]
\begin{center}
\epsfig{file=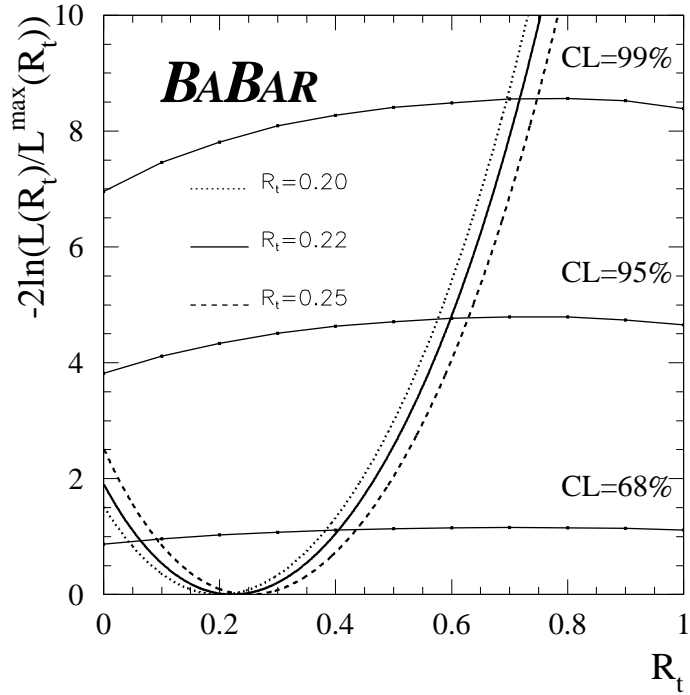,width=9cm}
\put(-105,265){{\large \bf PRELIMINARY}}
\caption{Likelihood ratio 
$L_{data}(R_t)=-2\ln({\cal{L}}(R_t)/{\cal{L}}(\bar{R_t}))$ 
for the data, where $\bar{R_t}$ is the value of $R_t$ that maximizes 
${\cal{L}}(R_t)$, and 68\%, 95\%, 99\% confidence level limits on $R_t$.
The systematic error on $R_t$ is indicated by the dotted and dashed lines.}
\label{cl}
\end{center}
\end{figure}

The dominant systematic uncertainty on the measured value of $R_t$
comes from the acceptance bias due to the transverse momentum 
threshold in the detection of soft pions. This bias ranges 
between $-0.048$ and $+0.004$, and depends on the yet unknown 
decay amplitudes. The central value of this interval was taken 
as a correction to the fitted $R_t^{sig}$, while its half width was assumed as 
a contribution to the systematic error. 
Other contributions come from the imperfect knowledge of the 
finite resolution on the measured value of
$\theta_{tr}$ (0.006), of the angular distribution of the
background (0.008) and of the purity of the signal sample (0.0003).  
All uncertainties were evaluated with Monte Carlo simulations.  The total systematic uncertainty on
$R_t$ was determined to be 0.031, giving the final corrected
result:
$$R_t = 0.22 \pm 0.18(stat) \pm 0.03(syst)$$

The likelihood function was also used to estimate approximate confidence 
intervals, taking into account the allowed physical region.
Fig.~\ref{cl} shows the dependence of 
$L_{data}(R_t)=-2\ln({\cal{L}}(R_t)/{\cal{L}}(\bar{R_t}))$ on $R_t$,
where $\bar{R_t}$ is the value of $R_t$ that maximizes ${\cal{L}}(R_t)$.
The lines corresponding to the 68\%, 95\%, 99\% confidence level 
limits on $R_t$ are also plotted. 
We can exclude values of $R_t$ greater than 0.63 at the 95\%
confidence level.

\section{Observation of the decays \Bztodstd and \Bpmtodstdstz}
\label{sec:Observation}
The decays \Bztodstdcc\ and \Bpmtodstdstz\ are studied following a method
largely similar to that described in Sec.~\ref{sec:Analysis}.  Here, only
those aspects of the analyses that differ significantly from that of the \Bztodstdst\ 
analysis are discussed in some detail.

For \Bztodstd , \Bz mesons are exclusively reconstructed by combining a 
\Dstarpm and a \Dmp candidate that are reconstructed in a number of
\Dstarpm and \Dmp decay modes.  For \Bpmtodstdstz\, the exclusive reconstruction combines
a \Dstarpm and a \Dstarz.  The kaon flavor of the \Dstarz is 
checked to make sure that a \Dstarp is paired only with a \Dstarzb and a \Dstarm is only 
paired with a \Dstarz.  The selection of \Dpm and \Dstarpm candidates, and the \KS and
\piz candidates that are used to compose them, is identical to that
described for the \Bztodstdst\ analysis.  

The decay modes of the $D$ and \Dstar used in these analyses are selected by an optimization
of \SsqovSpB based on Monte Carlo simulations.  \Dstarpm mesons are reconstructed
in their decays \Dstptopip and \Dstptopiz, and \Dstarz mesons are reconstructed in their
decays \Dstztopiz and \Dstztogam.  Modes used and their branching fractions are summarized
in Tables~\ref{dzbr} and \ref{dstarbr}.  As in the \Bztodstdst\ analysis, we construct
\chisqM variables that include all measured \Dstarpm, \Dstarz, and $D$ masses.  For \Bpmtodstdstz,
\chisqM contains 4 terms:

\begin{eqnarray*} \chisqM =&
   \left(\frac{m_D - m_{D_{PDG}}}{\sigma_{m_D}}\right)^2
 + \left(\frac{m_{\Db} - m_{\Db_{PDG}}}{\sigma_{m_{\Db}}}\right)^2 \\
  &\quad + \left(\frac{\Delta m_{\Dstar} - \Delta
   m_{\Dstar_{PDG}}}{\sigma_{\Delta m_{D^{*}}}}\right)^2
 + \left(\frac{\Delta m_{\Dstarz} - \Delta
   m_{\Dstarz_{PDG}}}{\sigma_{\Delta m_{D^{*0}}}}\right)^2
\end{eqnarray*}

For \Bztodstd , \chisqM contains 3 terms:

$$\chisqM =
   \left(\frac{m_D - m_{D_{PDG}}}{\sigma_{m_D}}\right)^2
 + \left(\frac{m_{D_{\Dstar}} - m_{D_{PDG}}}{\sigma_{m_{D_{\Dstar}}}}\right)^2
 + \left(\frac{\Delta m_{\Dstar} - \Delta
   m_{\Dstar_{PDG}}}{\sigma_{\Delta m_{D^{*}}}}\right)^2 $$

The major difference between these analyses and the \Bztodstdst\ analysis is that the \chisqM cut values
for these analyses are set individually for each submode instead of having a global
\chisqM value for all submodes, to better take into account the fact that the amount of background is 
quite different for each of the different submodes in these analyses.  The procedure used to
optimize the \chisqM value chooses the individual values of \chisqM that together
maximize the global \SsqovSpB.  
Submodes which have an optimal value of less than 2 times the number of degrees of freedom of the \chisqM
(8 for \Bpmtodstdstz, 6 for \Bztodstd) are rejected on the grounds that the resulting sensitivity is
poor, and that the tightness of the \chisqM value  makes such modes more susceptible to systematic errors in
measured yield.

\begin{figure}[p]
\begin{minipage}[t]{3.3in}
\begin{center}
\epsfig{file=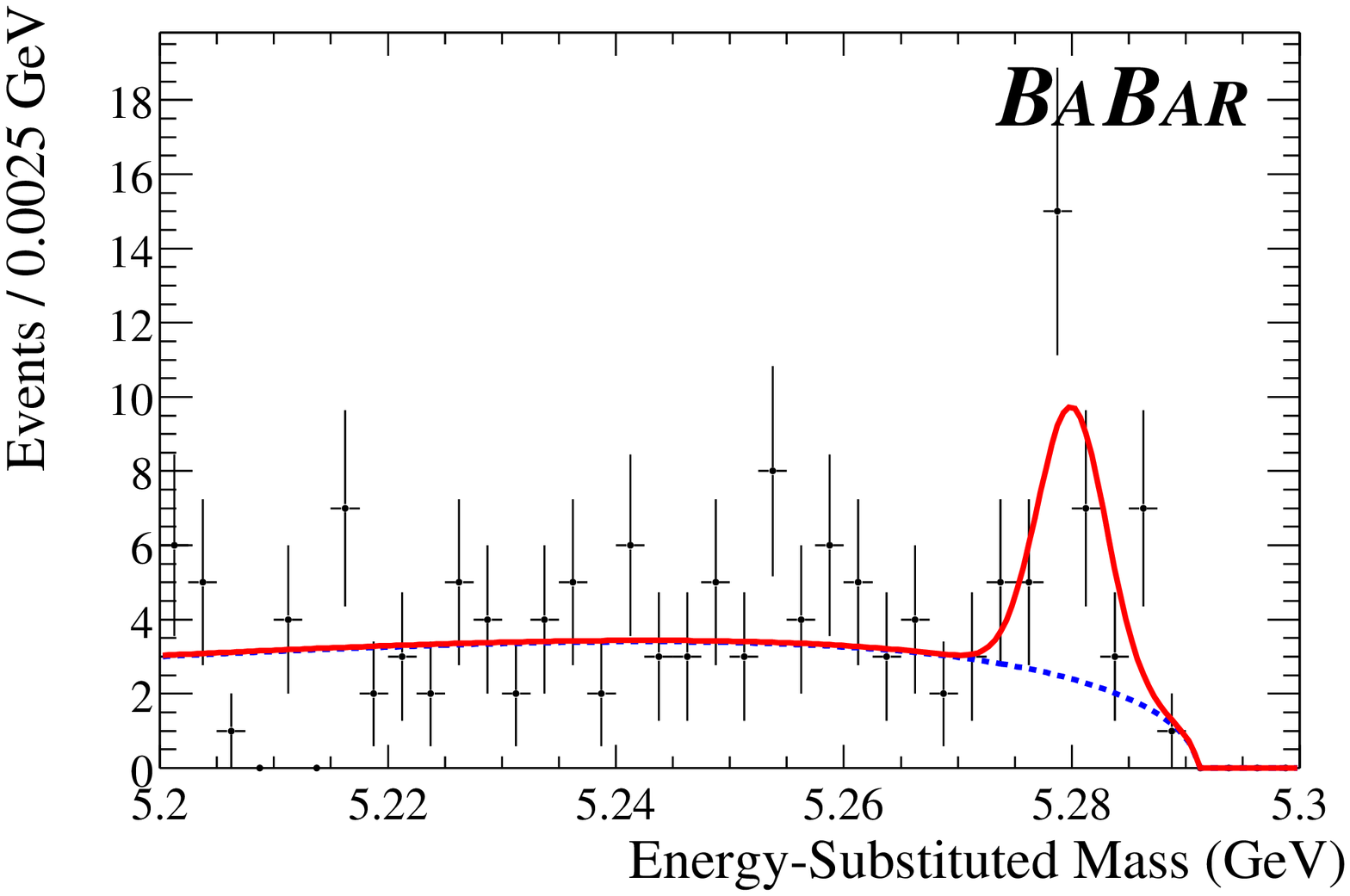,width=8.3cm}
\put(-190,155){{\large \bf PRELIMINARY}}
\end{center}
\end{minipage}
\begin{minipage}[t]{3.3in}
\begin{center}
\epsfig{file=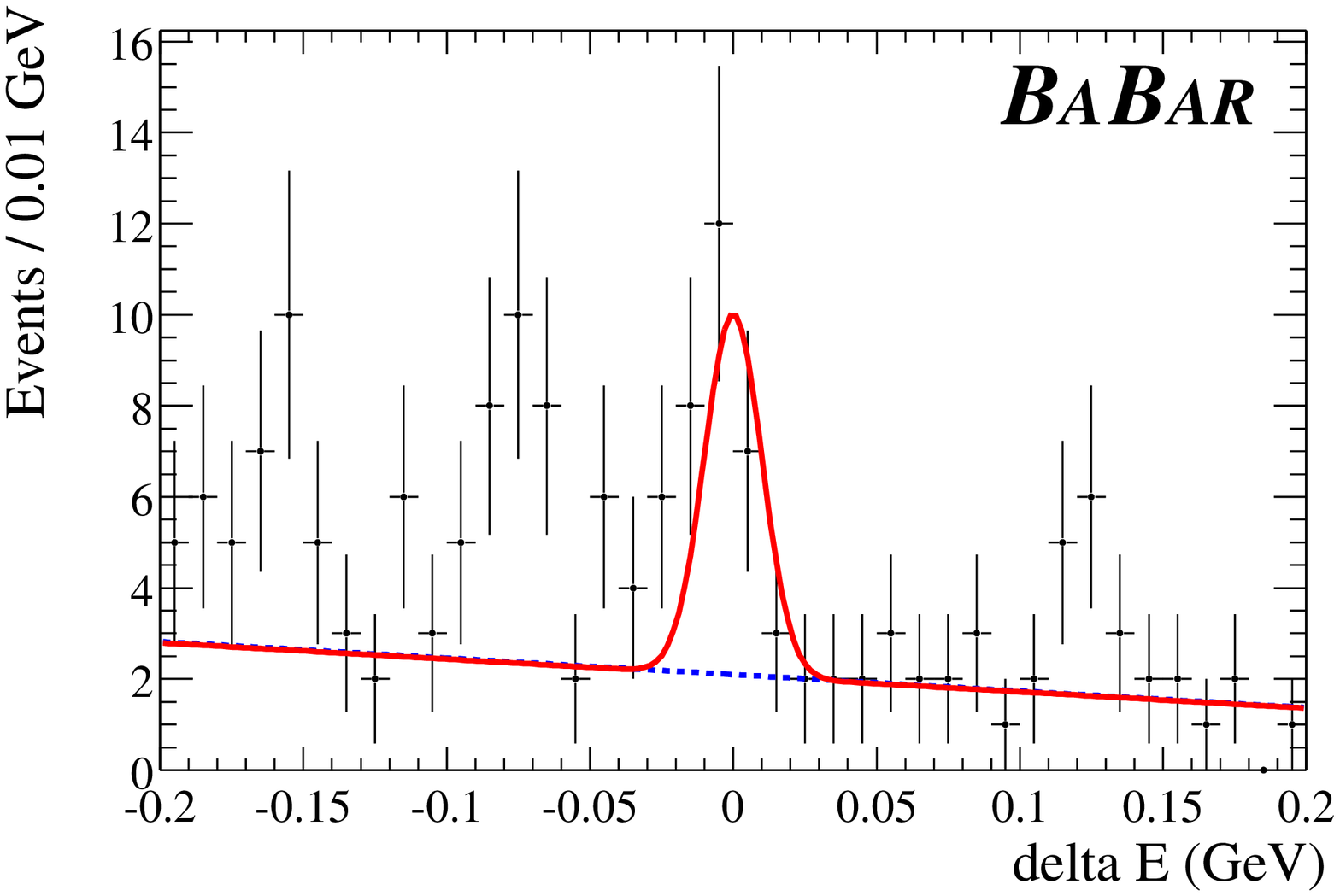,width=8.3cm}
\end{center}
\end{minipage}
\begin{center}
\epsfig{file=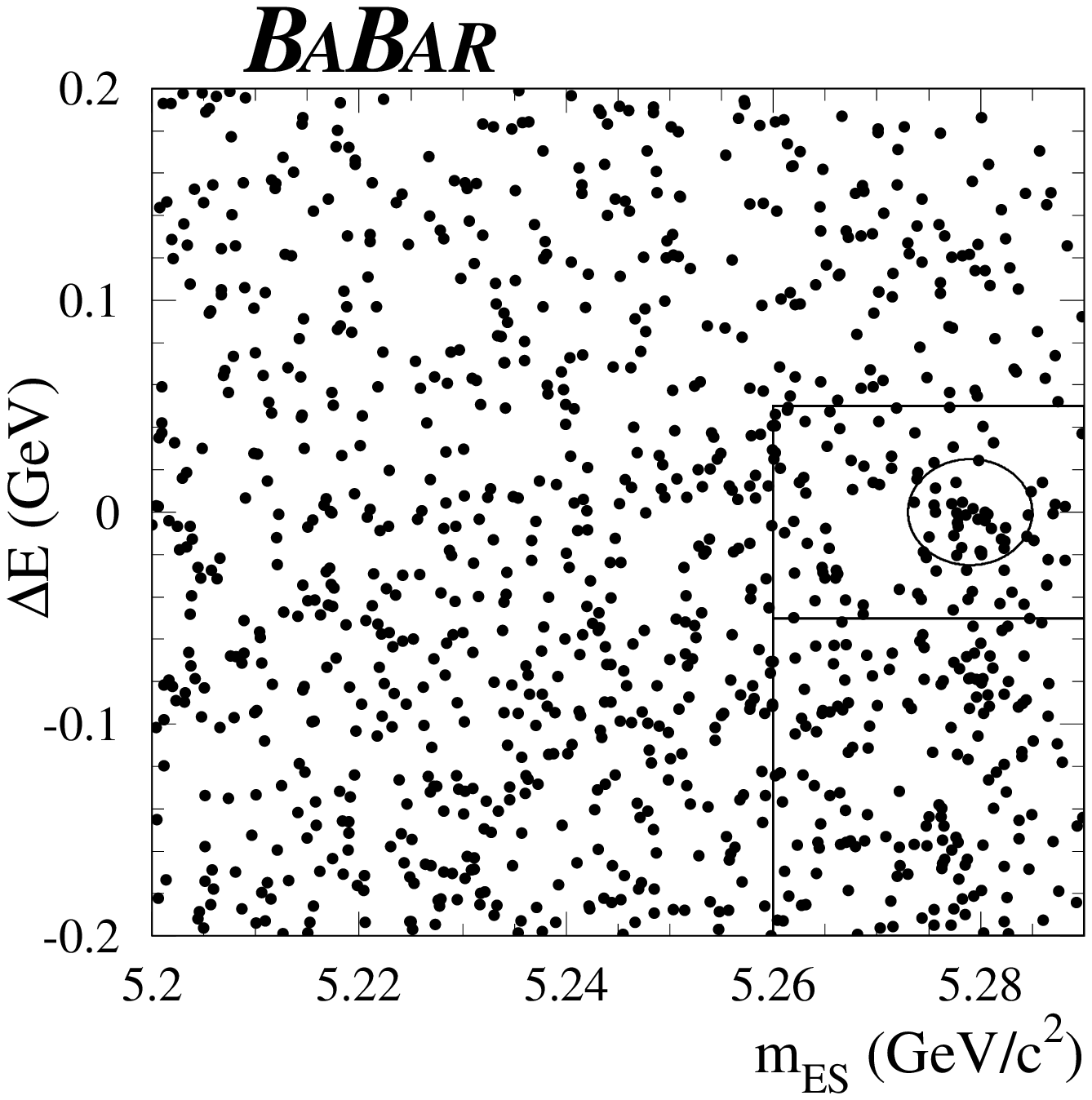,width=8.5cm}
\end{center}
\caption{
\label{PB0DsD}
      Top Left: \mes projection of the \Bztodstd\ event population, in the
      \DeltaEStd signal band ($-0.025 < \DeltaEStd < 0.025 \gev$). The  crosses
      are the data. The dashed line represents the extrapolation, to
      the \DeltaEStd signal band, of a two-dimensional background fit in the
      \DeltaEStd-\mes sidebands. The solid line is the sum of this background
      extrapolation and of a fitted, Gaussian-shaped signal centered on
      the $B$ mass.
      Top Right: \DeltaEStd projection of the \Bztodstd\ event population, in the
      \mes signal band ($5.273 < \mes < 5.285 \gevcc$). The crosses are the data. The dashed line
      represents the extrapolation, to the \mes signal band, of the
      above-mentioned background fit. The solid line is the sum of this
      background extrapolation, and of a gaussian-shaped signal
      centered on $\DeltaEStd = 0$. In the case of both the upper plots, the points in the feed-down region
      ($\DeltaEStd < -0.05 \gev$) are excluded from the fits.
      Bottom: Two-dimensional distribution of the \Bztodstd\ events in the
      \DeltaEStd \textit{vs.} \mes plane.  The small ellipse indicates the signal
      region, while the sideband region is everything that is outside the box that surrounds the signal region
      and also outside the (feed-down) box below the signal region.
}
\end{figure}

\begin{figure}[p]
\begin{minipage}[t]{3.3in}
\begin{center}
\epsfig{file=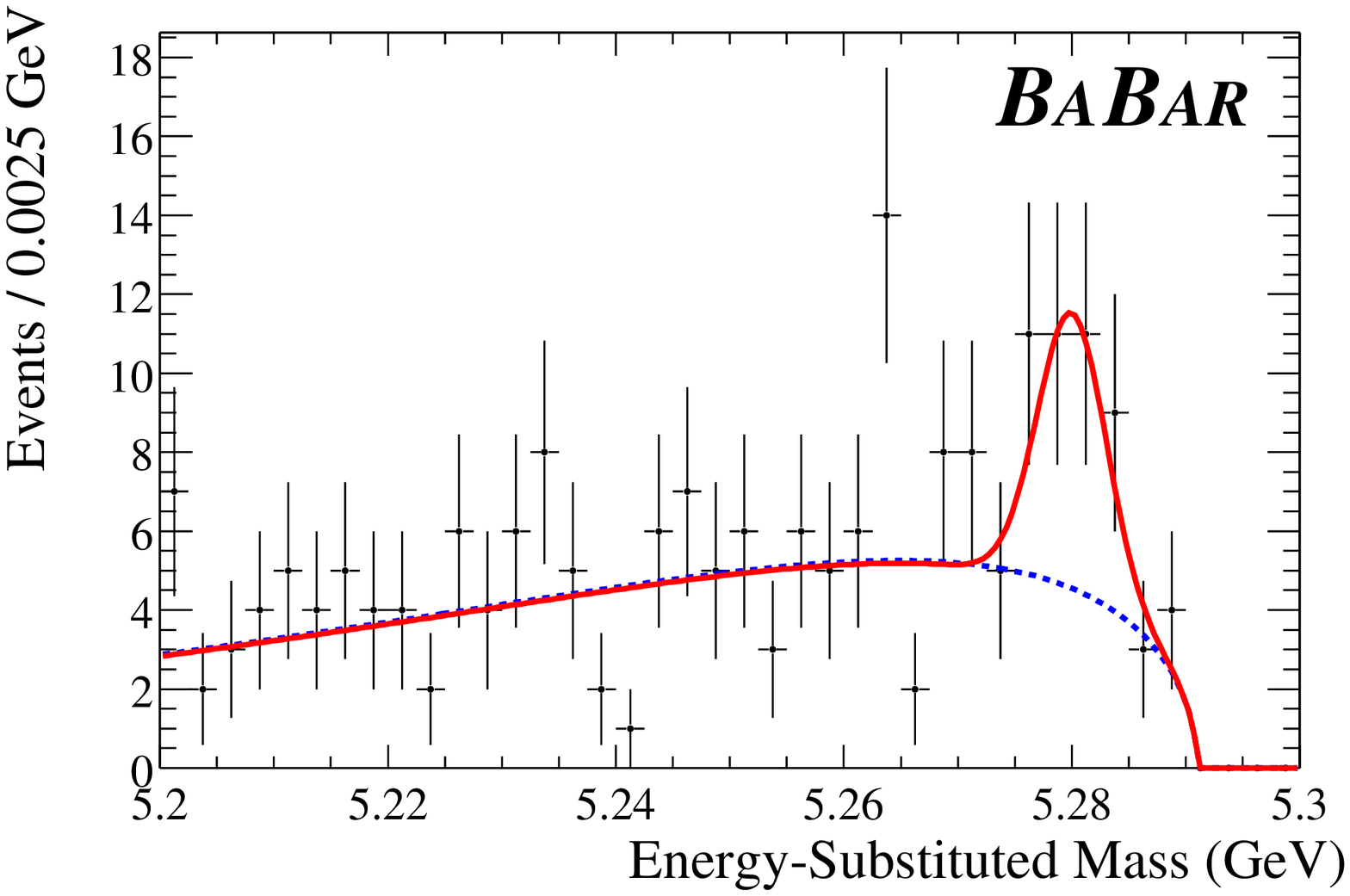,width=8.3cm}
\put(-190,155){{\large \bf PRELIMINARY}}
\end{center}
\end{minipage}
\begin{minipage}[t]{3.3in}
\begin{center}
\epsfig{file=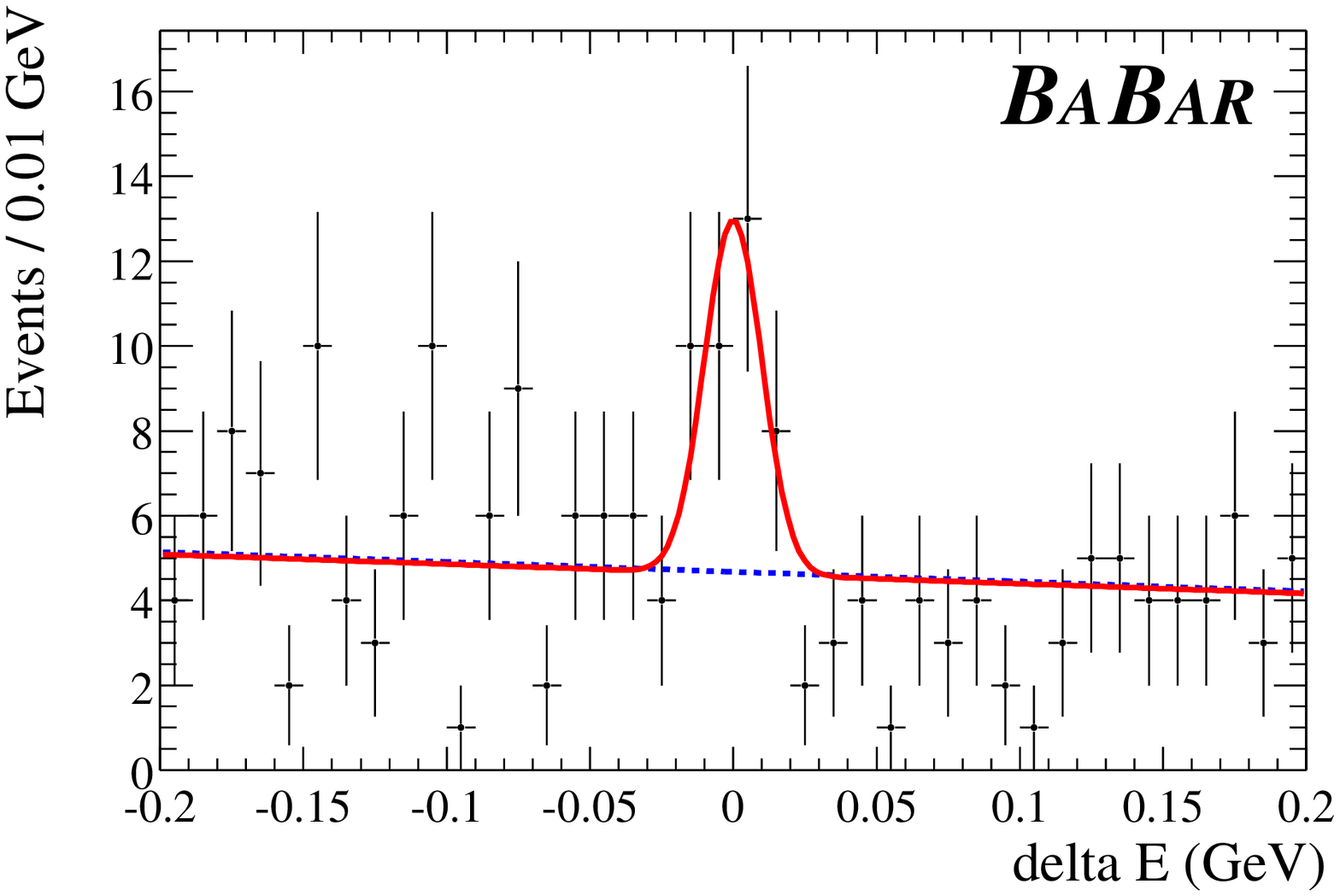,width=8.3cm}
\end{center}
\end{minipage}
\begin{center}
\epsfig{file=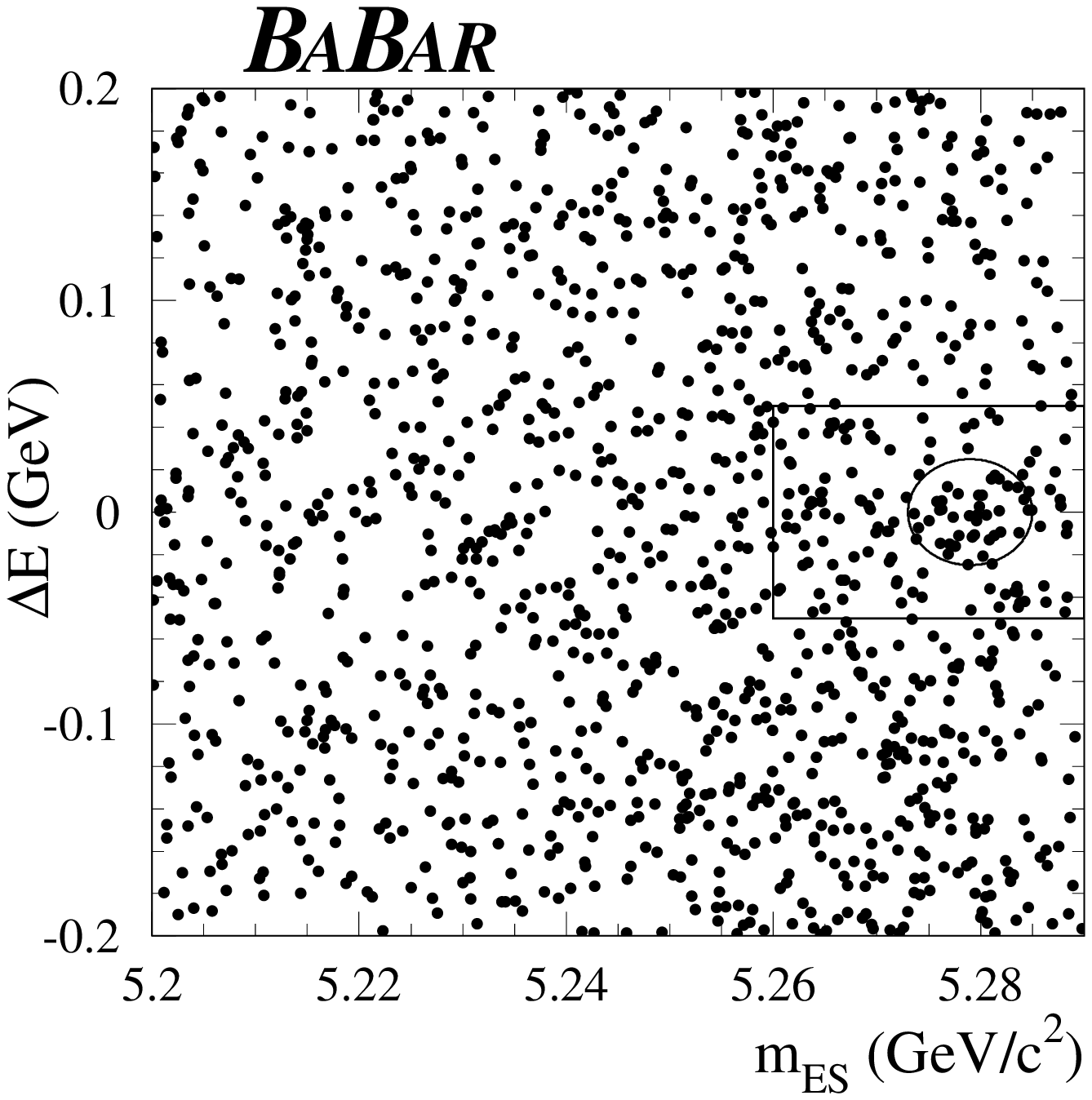,width=8.5cm}
\end{center}
%\put(-105,185){{\large \bf PRELIMINARY}}
\caption{
\label{PBchDsDsz}
      Top Left: \mes projection of the \Bpmtodstdstz\ event population, in the
      \DeltaEStd signal band ($-0.025 < \DeltaEStd < 0.025 \gev$). The crosses
	 are the data. The dashed line represents the extrapolation, to
      the \DeltaEStd signal band, of a two-dimensional background fit in the
      \DeltaEStd-\mes sidebands. The solid line is the sum of this background
      extrapolation and of a fitted, Gaussian-shaped signal centered on
      the $B$ mass.
      Top Right: \DeltaEStd projection of the \Bpmtodstdstz\ event population, in the
      \mes signal band ($5.273 < \mes < 5.285 \gevcc$). The crosses are the data. The dashed line
      represents the extrapolation, to the \mes signal band, of the
      above-mentioned background fit. The solid line is the sum of this
      background extrapolation, and of a Gaussian-shaped signal
      centered on $\DeltaEStd = 0$.
      Bottom: Two-dimensional distribution of the \Bpmtodstdstz\ events in the
      \DeltaEStd \textit{vs.} \mes plane.  The small ellipse indicates the signal
      region, while the sideband region is everything that is outside the box that surrounds the signal region.
}

\end{figure}

These values were tuned with samples of signal and generic \BB and \ccbar Monte Carlo where
the background distribution is taken from a sideband region.  For \Bpmtodstdstz , the sideband region is the same as
that for \Bztodstdst.  For \Bztodstd , the region where decays
(such as \Bztodstdst) can feed down into the \DeltaEStd-\mes plane must be eliminated from the sideband.
As \Bztodstdst contains a reconstructed \Dstarpm and $D$, it is only separated from \Bztodstd events due to the
missing energy of the slow pion from the second \Dstarpm.  This missing energy manifests itelf as a negative shift in
\DeltaEStd.  These events accumulate in the area below the \Bztodstd signal region, and in order to remove them,
 the region defined by:
$$ \DeltaEStd < -50\mev $$
$$ \mes > 5.26\gevcc $$ 
is removed.  The \mes\ and \DeltaEStd
distributions for events reconstructed in the channels
\Bztodstd\ and \Bpmtodstdstz\ are shown in Fig.~\ref{PB0DsD} and 
in Fig.~\ref{PBchDsDsz}. 

For the \Bztodstd\ channel we reconstruct a total of 31 events, of which
$10.5 \pm 1.7(stat)$ are background.
The probability that the visible signal is a statistical fluctuation
of the background is $9.7\times 10^{-7}$ ($>4.3\sigma$). 
As pointed out in the introduction, 
this channel is a \CP\ conjugate state that can
be used for \stwob measurements.
For \Bpmtodstdstz we reconstruct a total of 39 events, of which
$20.3 \pm 0.5(stat)$ are background.
The probability that the visible signal is a statistical fluctuation
of the background is $2.9\times 10^{-6}$ ($>4.1\sigma$). 
This channel is useful for calibrations and tests of the \CP\ fitting
procedures.  

\section{Summary}
\label{sec:Summary}
%Succinctly summarize the result and conclusions.
Using data collected by the \babar\ experiment during 1999-2000, we
have observed a signal of $31.8 \pm 6.2(stat) \pm 0.4(syst)$ events in
the decay \Bztodstdst.  Our measurement of the branching ratio is
$$\BRbztodstdst$$
From the transversity angular distribution of these events,
the measured fraction of the component with odd \CP
parity is 
$$R_{t} = 0.22 \pm 0.18(stat) \pm 0.03(syst)$$
Finally, signals are also observed in the decay modes
\Bztodstd and \Bpmtodstdstz.

\section{Acknowledgments}
\label{sec:Acknowledgments}

% Specific acknowledgments for this paper; remove if not needed.
%The authors wish to thank Prof.\ A.\ Cleverguy for his help
%with the theoretical interpretation of these results.

% Standard acknowledgments paragraph; must always be included.
We are grateful for the 
extraordinary contributions of our \pep2\ colleagues in
achieving the excellent luminosity and machine conditions
that have made this work possible.
The collaborating institutions wish to thank 
SLAC for its support and the kind hospitality extended to them. 
This work is supported by the
US Department of Energy
and National Science Foundation, the
Natural Sciences and Engineering Research Council (Canada),
Institute of High Energy Physics (China), the
Commissariat \`a l'Energie Atomique and
Institut National de Physique Nucl\'eaire et de Physique des Particules
(France), the
Bundesministerium f\"ur Bildung und Forschung
(Germany), the
Istituto Nazionale di Fisica Nucleare (Italy),
the Research Council of Norway, the
Ministry of Science and Technology of the Russian Federation, and the
Particle Physics and Astronomy Research Council (United Kingdom). 
Individuals have received support from the Swiss 
National Science Foundation, the A. P. Sloan Foundation, 
the Research Corporation,
and the Alexander von Humboldt Foundation.


\begin{thebibliography}{99}

\bibitem{ref:cc}
Charge-conjugate states are implied throughout this paper and the symbol
$D^{(*)}$ refers to either $D$ or $D^*$.

\bibitem{GWF} Y.~Grossman and M.~Worah, \plb {\bf 395}, 241 (1997);\\
R.~Fleischer, \ijmp A {\bf 12}, 2459 (1997).


\bibitem{ref:dunietz}
I.~Dunietz {\it et al.}, \jprd {\bf 43}, 2193 (1991).

\bibitem{cleoprd62} E.~Lipeles {\it et al.}, \jprd {\bf 62}, 032005 (2000).

\bibitem{alephepj98} R.~Barate {\it et al.}, \epj {\bf C4}, 387 (1998).

\bibitem{ref:babar}
The \babar\ Collaboration, B.\ Aubert {\em et al.},
SLAC-PUB-8596, hep-ex/0105044, to appear in Nucl.\ Instrum.\ Methods.

\bibitem{ref:pepii}
PEP-II Conceptual Design Report, SLAC-R-418 (1993).

\bibitem{ref:fox}
G.~C.~Fox and S.~Wolfram, \jprl {\bf 41}, 1581 (1978).

\bibitem{pdg} Particle Data Group, D.~E.~Groom {\it et al.},
\epjc {\bf 15}, 1 (2000).

\bibitem{ref:argus}
ARGUS Collaboration, H.~Albrecht {\it et al.}, \jpl {\bf B185}, 218 (1987).

\bibitem{transamplitudes} P. H. Harrison and H. R. Quinn, eds. ``The \babar\ 
Physics Book'', SLAC-R-504 (1998).



%\bibitem{ref:cc}
%Charge-conjugate states are implied throughout.

%\bibitem{BAD134} C.~Hearty, ``Measurement of the Number of \FourS Mesons 
%Produced in Run 1 (B Counting)'', \babar\ Analysis Document 134,
%(2001).

%\bibitem{BAD150} BReco AWG, ``Exclusive Reconstruction of Hadronic $B$ 
%Decays to Open Charm'', \babar\ Analysis Document 150, (2001).

%\bibitem{lumi}
%{\tt \small http://www.slac.stanford.edu/BFROOT/www/Physics/BaBarData/GoodRuns/dataSets.html}

%\bibitem{KaonID} 
%{\tt \small http://www.slac.stanford.edu/BFROOT/www/Physics/Tools/BetaTools/MicroKilling.html} \\
%\bibitem{sysKaonID} 
%private communication Aaron Roodman


%\bibitem{TrkEff}
%E. Varnes`` Measurement of the Tracking Efficiency Using 3+1 tau Events '', \babar\  Analysis Document 87, (2000).\\
%{\tt \small http://babar-hn.slac.stanford.edu:5090/HyperNews/get/physAnal/1037.html \\
%http://babar-hn.slac.stanford.edu:5090/HyperNews/get/physAnal/1024.html \\
%http://www.slac.stanford.edu/BFROOT/www/Physics/\\ 
%TrackEfficTaskForce/Recipe/TrackingEfficiencies.html}

%\bibitem{KSEff}
%{\tt \small http://babar-hn.slac.stanford.edu:5090/HyperNews/get/physAnal/1029.html}
%{\tt \small http://babar-hn.slac.stanford.edu:5090/HyperNews/get/physAnal/1042.html}

%\bibitem{PiZeroEff}
%{\tt \small http://babar-hn.slac.stanford.edu:5090/HyperNews/get/physAnal/1028.html}

\end{thebibliography}
\end{document}